\documentclass[%
 reprint,
superscriptaddress,
 amsmath,amssymb,
 aps,
]{revtex4-2}

\usepackage{graphicx}
\usepackage{dcolumn}
\usepackage{bm}

\begin{document}


\title{Adaptive control in dynamical systems using reservoir computing}

\author{Swarnendu Mandal}
\affiliation{%
 International Research Center for Neurointelligence (WPI-IRCN), The University of Tokyo, Tokyo, Japan
}%

\author{Swati Chauhan}%
\affiliation{Department of Physics, Central University of Rajasthan, Ajmer, Rajasthan, India}%

\author{Umesh Kumar Verma}%
\affiliation{Department of Physics, Central University of Rajasthan, Ajmer, Rajasthan, India}%

\author{Manish Dev Shrimali}%
\affiliation{Department of Physics, Central University of Rajasthan, Ajmer, Rajasthan, India}%

\author{Kazuyuki Aihara}
\affiliation{%
 International Research Center for Neurointelligence (WPI-IRCN), The University of Tokyo, Tokyo, Japan
}%

\date{\today}

\begin{abstract}
We demonstrate a data-driven technique for adaptive control in dynamical systems that exploits the reservoir computing method. We show that a reservoir computer can be trained to predict a system parameter from the time series data. Subsequently, a control signal based on the predicted parameter can be used as feedback to the dynamical system to lead it to a target state. Our results show that the dynamical system can be controlled throughout a wide range of attractor types. One set of training data consisting of only a few time series corresponding to the known parameter values enables our scheme to control a dynamical system to an arbitrary 
 target attractor starting from any other initial attractor. In addition to numerical results, we implement our scheme in real-world systems like a R\"{o}ssler system realized in an electronic circuit to demonstrate the effectiveness of our approach.

\end{abstract}

\maketitle


\section{\label{sec:intro}Introduction}

Adaptive control refers to the mechanism in which a system modifies its properties to achieve a specific target state. The concept of adaptability spans a wide range of disciplines, from power grids~\cite{Khalid2021,Berner2021} that need to dynamically adjust their load to machine learning systems~\cite{Sutton1992}, where `learning' is synonymous with adaptation to new data. In machine learning, adaptability is fundamentally linked to the concept of learning, as systems continuously evolve based on the data or feedback they receive~\cite{Sutton1998}. Reinforcement learning exemplifies adaptability by employing feedback mechanisms to refine decision-making policies over time. In addition, unsupervised learning techniques, such as clustering and dimensionality reduction, allow systems to adaptively uncover patterns and structures in high-dimensional data, enabling their application to various tasks. Beyond engineering systems, adaptivity is also inherent in natural systems. For example, fireflies exhibit frequency adaptation in their synchronized flashing, a behavior governed by feedback interactions within the group~\cite{Mirollo1990}. In neuroscience, synaptic plasticity, a key feature of neuronal networks, allows adaptive learning and memory formation through mechanisms such as long-term potentiation and long-term depression~\cite{Bliss1973,Berner2019}.

A control mechanism is supposed to tune the available system parameter to lead the system to or maintain it in a target state, depending on the current measurement of the system dynamics. In 1990, Huberman \& Lumer \cite{huberman1990dynamics} proposed a control scheme in which a system parameter is adjusted based on the difference between the target dynamics and the current dynamics of the system. Along the same lines of exploration, the mechanism has been applied to a high-dimensional multi-parameter system to target periodic dynamics \cite{sinha1990adaptive} or even a chaotic attractor \cite{ramaswamy1998targeting}. More experimentally viable control mechanisms are given by OGY \cite{ott1989controlling} and Pyragas \cite{pyragas1992continuous} where {\em a priori} analytical information of the system dynamics is not required. In particular, these techniques are specifically useful for controlling chaos by stabilizing unstable orbits present in the chaotic dynamics. Whereas the Huberman \& Lumer approach can target any particular dynamics in the parameter space. The only limitation is that it requires the information of the mathematical model of the system. In this article, we utilize a machine learning technique to implement the control scheme in a model-free way. Specifically, we use a reservoir computer based on echo-state network (ESN) architecture for this purpose. In an earlier study \cite{haluszczynski2021controlling}, a reservoir computer is shown to be successful in targeting arbitrary attractors of a dynamical system. However, this approach is useful only for creating a digital twin of the target dynamics, not to control a physical system exactly. In recent developments, next-generation reservoir computers (NG-RC) \cite{kent2024controlling} and deep echo-State networks (dESN) \cite{canaday2021model} are employed to provide efficient techniques for chaos control based on the OGY and Pyragas approach.

In recent years, reservoir computing has been a great machine learning tool for applications in the field of dynamical systems due to its exceptional capability in processing temporal information. Among its numerous applications in time series prediction \cite{jaeger2004harnessing}, inferring unmeasured variables \cite{lu2017reservoir}, pattern recognition \cite{tanaka2022simulation}, anomaly detection \cite{kato2024reconstructive}, and many more \cite{pathak2017using,pathak2018model,lu2018attractor,mandal2023learning,restrepo2023suppressing,tang2020introduction,yan2024emerging}, one of the most important ones is to approximate the global behavior of a dynamical system learning from a few local examples for a model-free system. This refers to a reservoir computer's ability to accurately predict several transition and bifurcation points with varying system parameters, while it was trained by only local dynamics in the parameter space \cite{kim2021teaching,kong2021machine}. Not only can it reproduce the bifurcation diagram, but it is also capable of inferring the presence of an unknown attractor in a multistable system \cite{roy2022model}. A parameter-aware reservoir computer is employed to perform these kinds of tasks, in which a constant input channel is introduced to feed the information of the system parameters in addition to the usual state variables input. After training, the reservoir computer is expected to produce the system dynamics for a range of system parameters. In this article, we exploit an inverse architecture of a parameter-aware reservoir computer. The system parameters are used as labels to train the model to estimate the parameter values from the given time series as input. The scheme is to employ a reservoir computer to estimate parameter values using time series data in real-time. Then, a control mechanism based on the predicted parameter is used to drive the system to and maintain it in a particular target state.

\section{\label{sec:prob}Problem Setting}

Consider an $n$-dimensional dynamical system whose dynamics is described by 

\begin{equation}\label{eq:sys_gen}
    \bm{\dot{X}} \equiv \frac{d\bm{X}}{dt} = \mathcal{F}(\bm{X};\Lambda;t),
\end{equation}
where, $\bm{X} \equiv [X_1,X_2,....,X_n]^T$ is the vector of state variables of the system. $\Lambda \equiv (\lambda_1,\lambda_2,.......,\lambda_m)$ are the system parameters which control the system dynamics.

The popular way of adaptive control is to introduce additional dynamics of the system parameters as

\begin{equation}
    \frac{d\Lambda}{dt} = \epsilon(\mathcal{S}^* - \mathcal{S}),
\end{equation}
where, $\mathcal{S}^*$ corresponds to the target value of $\mathcal{S}$. $\mathcal{S}$ can be state variables or some functions of them, representing a property of the dynamics. For example, to target a steady state by adaptive control $\mathcal{S}$ can be only state variables $\bm{X}$, whereas, chaotic dynamics is targeted using Lyapunov exponents functioning as $\mathcal{S}$. $\epsilon$ is termed as `stiffness' of control which decides the rate of control.

In this article, we propose a control scheme with an additional physical input,  added to the system to adjust parameter values of a physical dynamical system. The external input is controlled by a reservoir computer, based on the measurement of the current state of the system. Consider a physical system with the external control input as

\begin{equation}\label{eq:sys_control}
    \bm{\dot{X}} \equiv \frac{d\bm{X}}{dt} = \mathcal{F}(\bm{X};\Lambda_c;t).
\end{equation}

Here, The control input $I$ is added with the actual system parameter $\Lambda$ resulting an effective parameter, $\Lambda_c = \Lambda + I$. This external control input $I \equiv I(\Delta\Lambda)$ is a function of $\Delta\Lambda$, the difference between the target parameter value ($\Lambda^*$) and the current value ($\Lambda$).

\section{\label{sec:numsim}Numerical Simulation}

\subsection{\label{ssec:RC}Parameter Estimation}

The first step in the proposed scheme of model-free control of the dynamical systems is a real-time estimation of system parameters exploiting a reservoir computer (RC).  We employ an echo-state network model for this purpose. The input-driven reservoir dynamics is given by

\begin{equation}\label{eq:st_upd}
    \bm{r}(t+1) = (1-\alpha)\bm{r}(t) + \alpha~\mathrm{tanh}[\mathcal{W}^{res}\cdot \bm{r}(t) + \mathcal{W}^{in} \cdot \bm{u}(t)],
\end{equation}
where $\bm{r}(t)$ is the randomly initialized reservoir state and $\bm{u}(t)$ is the input vector containing the state variables at time instant $t$.
The reservoir network is defined by $\mathcal{W}^{res}$ and $\mathcal{W}^{in}$ is the input connection matrix. The weights of the input connection matrix $\mathcal{W}^{in}$ are randomly selected from range [-$\sigma$,$\sigma$].
The hyper-parameters for the setup is given in Table \ref{tab:hp}.

The reservoir states $\bm{r}(t)$ and the output $\bm{v}(t)$ is linked by a linear relation 

\begin{equation}
    \bm{v}(t) = \mathcal{W}^{out} \cdot \bm{r}(t),
\end{equation}
where $\mathcal{W}^{out}$ is the output connection matrix of ESN architecture. $\mathcal{W}^{out}$ is evaluated using the training dataset by Ridge regression. If $\mathcal{R}$ is the reservoir state matrix generated using training input $\tilde{u}(t)$ and $\mathcal{U}$ is the teacher matrix containing output labels $\tilde{v}(t)$ corresponding to $\tilde{u}(t)$, then

\begin{equation}
    \mathcal{W}^{out}=\mathcal{U}\mathcal{R}^T(\mathcal{R}\mathcal{R}^T + \beta I)^{-1}.
\end{equation}
$\beta$ is another hyper-parameter of the setup listed in Table \ref{tab:hp}.

To configure the ESN for parameter estimation from time series data, the input contains state variables of the system, i.e $\bm{u}(t) \equiv \bm{X}(t)$. The output contains the information of the parameter set. In training phase, the output is prepared by repeating the parameter values for all time steps as $[\bm{v}(t_1),\bm{v}(t_2), \bm{v}(t_3),........ ] = [\Lambda, \Lambda, \Lambda, ..........]$. During the testing phase, the predicted parameter values are estimated by averaging the output of ESN over time.

\begin{table}[h]
\centering
\begin{tabular}{lcc}
\hline
Hyperparameters & Symbols & Typical values \\
\hline
\hline
Leaking rate    &     $\alpha$    &        $0.5$         \\
\hline
Reservoir size  &     $N_{res}$   &       $1000$       \\
\hline
Spectral radius of $\mathcal{W}$  &  $\rho$  &  $0.5$   \\
\hline
Scaling factor of $\mathcal{W}_{in}$ & $\sigma$ &  $0.1$   \\
\hline
Regularization parameter &  $\beta$    &      $1\times 10^{-6}$      \\
\hline
\end{tabular}
\caption{\label{tab:hp}Hyper-parameters for the echo-state network}
\end{table}

\begin{figure}
    \centering
    \includegraphics[width=\linewidth]{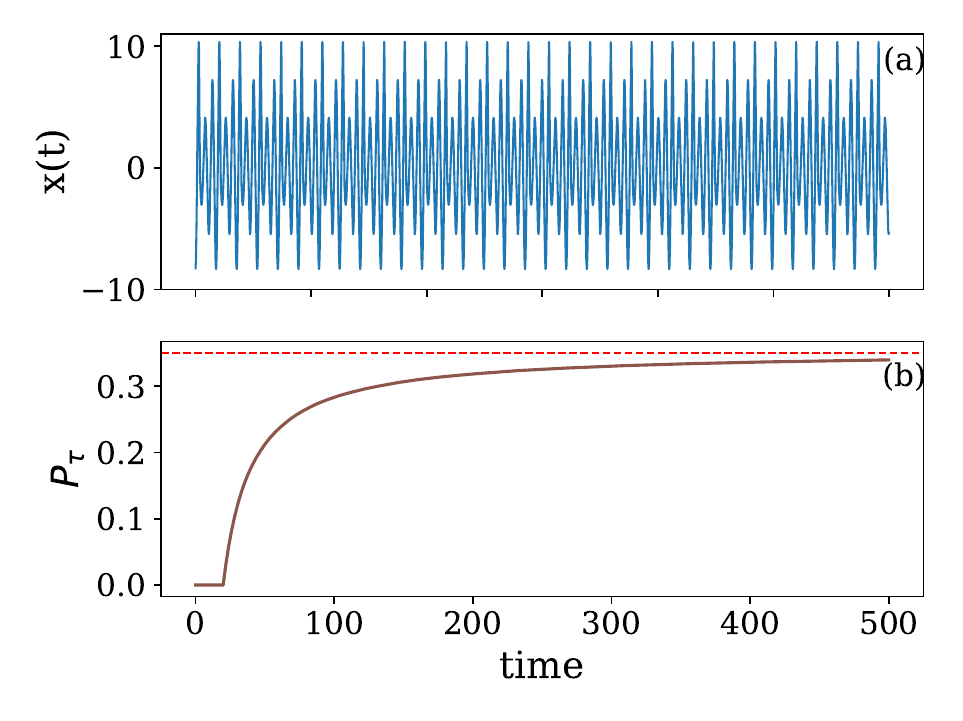}
    \caption{Parameter prediction of the R\"{o}ssler system. (a) depicts the input time series and, (b) is the plot of cumulative moving average of the output of ESN. The dashed red line represents the actual value of the parameter.}
    \label{fig:ros_par}
\end{figure}

For real-time prediction, we consider a 2nd order-averaging method of the ESN output.  We evaluate the cumulative average of the moving average of a specified window on the temporal axis. The Cumulative moving average ($P_\tau$) at any discrete time step $\tau$ can be defined as 

\begin{align}
    P_\tau = \frac{(\tau-1)P_{\tau-1}}{\tau} + \frac{C_\tau}{\tau},
\end{align}
where $C_\tau$ is the moving average determined by averaging the last $m$ steps of the predicted output. $m$ is the window length for the moving average. The result for parameter estimation is depicted in Fig. \ref{fig:ros_par}.

\subsection{\label{ssec:ross}Control in R\"{o}ssler System}

\begin{figure}
    \centering
    \includegraphics[width=\linewidth]{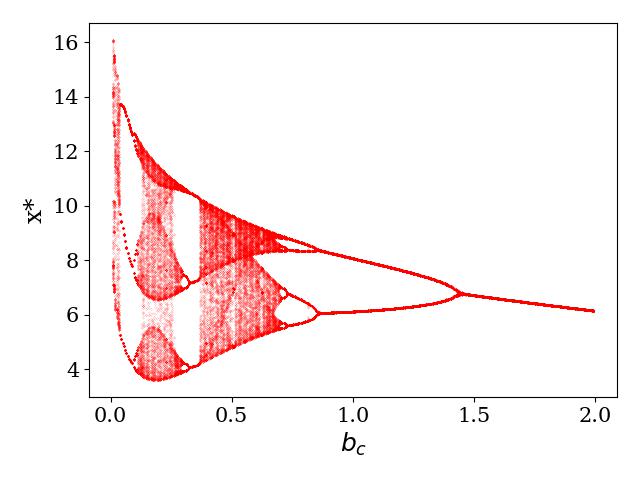}
    \caption{Bifurcation diagram of the R\"{o}ssler system (\ref{eq:rossler}) with varying $b_c$, where $a = 0.2$ and $c = 5.7$ are fixed. $x*$ denotes the maxima of the state variable $x(t)$.}
    \label{fig:bif_ros}
\end{figure}

Suppose that the R\"{o}ssler system is our control system given by

\begin{align}\label{eq:rossler}
\dot{x} &= - y - z,  \nonumber \\
\dot{y} &= x + ay ,  \\
\dot{z} &= b_c + x(z-c).  \nonumber  
\end{align}
Here, $b_c = b + b'$ is the parameter subjected to control. $b$ is the actual parameter of the system, vulnerable to unwanted perturbation and $b'$ corresponds to the RC driven control input, meant to stabilize the system against external perturbations and lead to stable target dynamics. The dynamics with varying $b_c$ of system (\ref{eq:rossler}) is given in figure \ref{fig:bif_ros}. If the target dynamics corresponds to parameter value $b_c = b^*$, the evolution of the control input can be described as

\begin{equation}\label{eq:control_ross}
    \frac{db'}{dt} = \varepsilon_{ross}(b^* - b_p),
\end{equation}
where $\varepsilon_{ross}$ is the stiffness of control as described by Huberman and Lumer \cite{huberman1990dynamics}. $b_p$ is the predicted value of $b_c$. For a perfectly trained machine, $b_p = b_c$.

\begin{figure*}
    \centering
    \includegraphics[width=\textwidth]{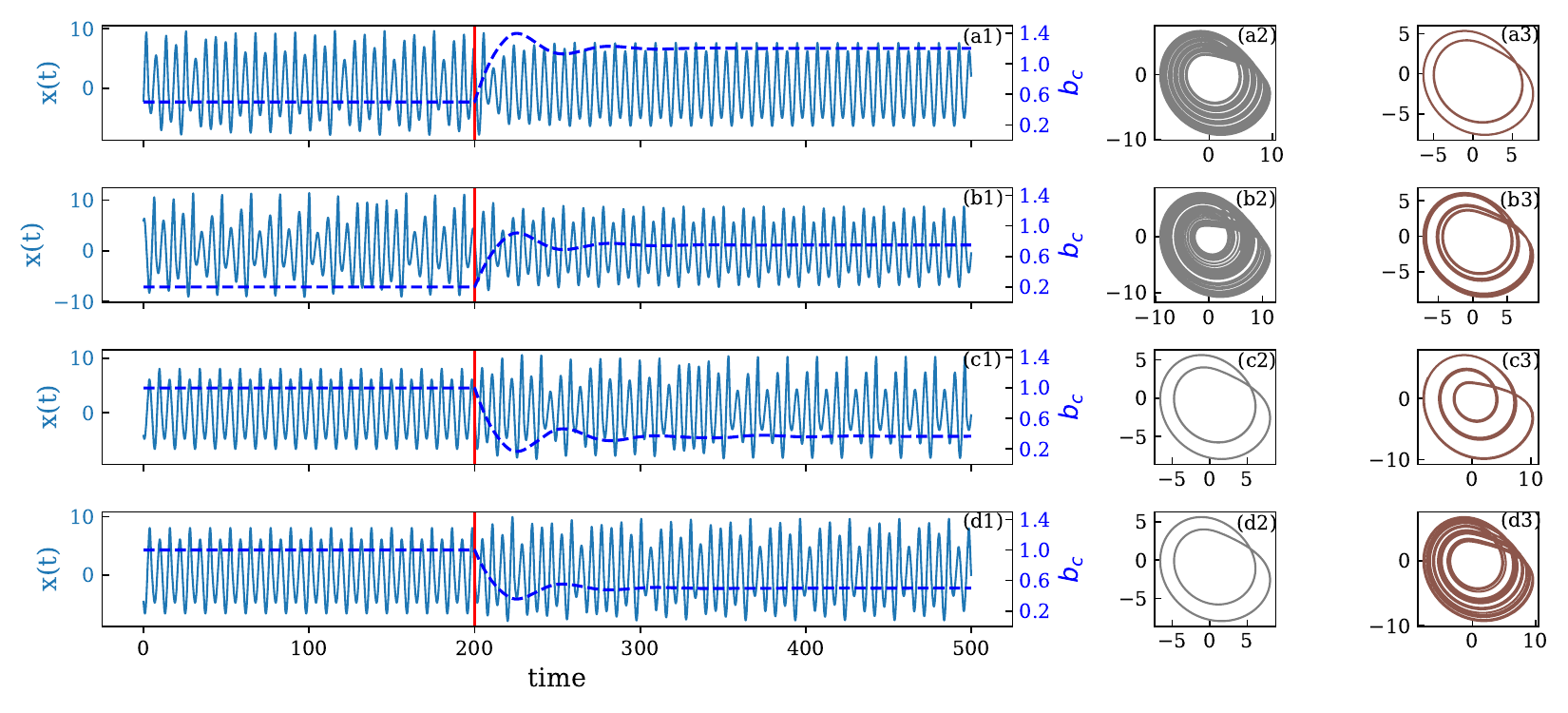} 
    \caption{Numerical results for control with parameter $b_c$ of the R\"{o}ssler system (\ref{eq:rossler}). The first column shows the time series of the system with control mechanism in action (blue lines) and the variation of the control parameter $b_c$ (dashed dark blue lines). The vertical red lines denotes the start of application of the control signal. The subsequent two columns represents corresponding initial and target attractors respectively in the $x-y$ plane. For all four rows, the corresponding initial and target values of parameter $b_c$ are given by $(0.5,1.2),~ (0.2,0.75),~(1.0,0.35)$, and $(1.0,0.5)$ respectively.}
    \label{fig:control_ross}
\end{figure*}

The role of the RC is to estimate the current value of $b_c$ based on the measurement of state variables. Then $b'$ is evaluated based on a given value of $b^*$ using equation (\ref{eq:control_ross}). Note that the value of $b^*$ can also be specified by providing the target dynamics itself, as a trained RC model can estimate parameter value from time series data. Results of the numerical simulation of the control mechanism is presented in Figure \ref{fig:control_ross} where the parameter-aware reservoir is trained with dynamics at $b_c = 0.6,~0.8,$ and $1.0$. After training, the value of $b_p$ is evaluated real-time by calculating the moving average of the ESN output over finite window length. For the results presented in this article, the window length is considered $200$ while the time step length of the input time series is $\Delta t = 0.1$. The value of stiffness of control is $\varepsilon_{ross} = 0.01$ for the numerical simulations.  

\subsection{\label{ssec:logm}Control in Logistic Map}

\begin{figure}[h!]
    \centering
    \includegraphics[width=\linewidth]{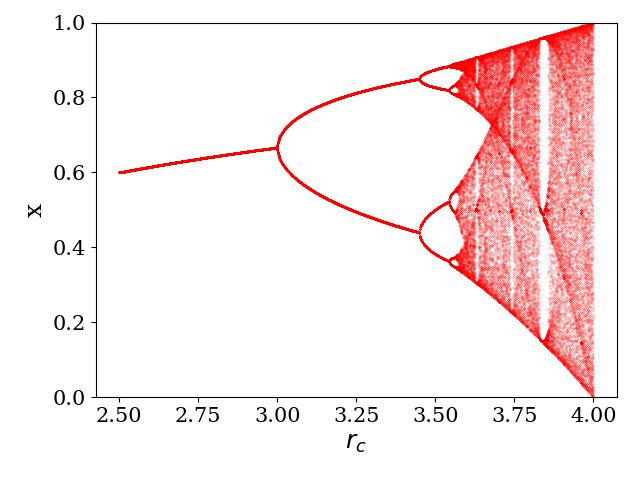}
    \caption{Bifurcation diagram of the Logistic map (\ref{eq:logm}) with varying $r_c$}.
    \label{fig:bif_logm}
\end{figure}

We consider the logistic map for applying the control mechanism on a discrete-time system. The equation of the map is given by

\begin{equation}\label{eq:logm}
    x_{n+1} = r_cx_n(1-x_n),
\end{equation}
where $r_c$ is the system parameter which is subject to control. The change in dynamics of the logistic map can be captured in bifurcation diagram presented in Figure \ref{fig:bif_logm}.

\begin{figure*}
    \centering
    \includegraphics[width=\textwidth]{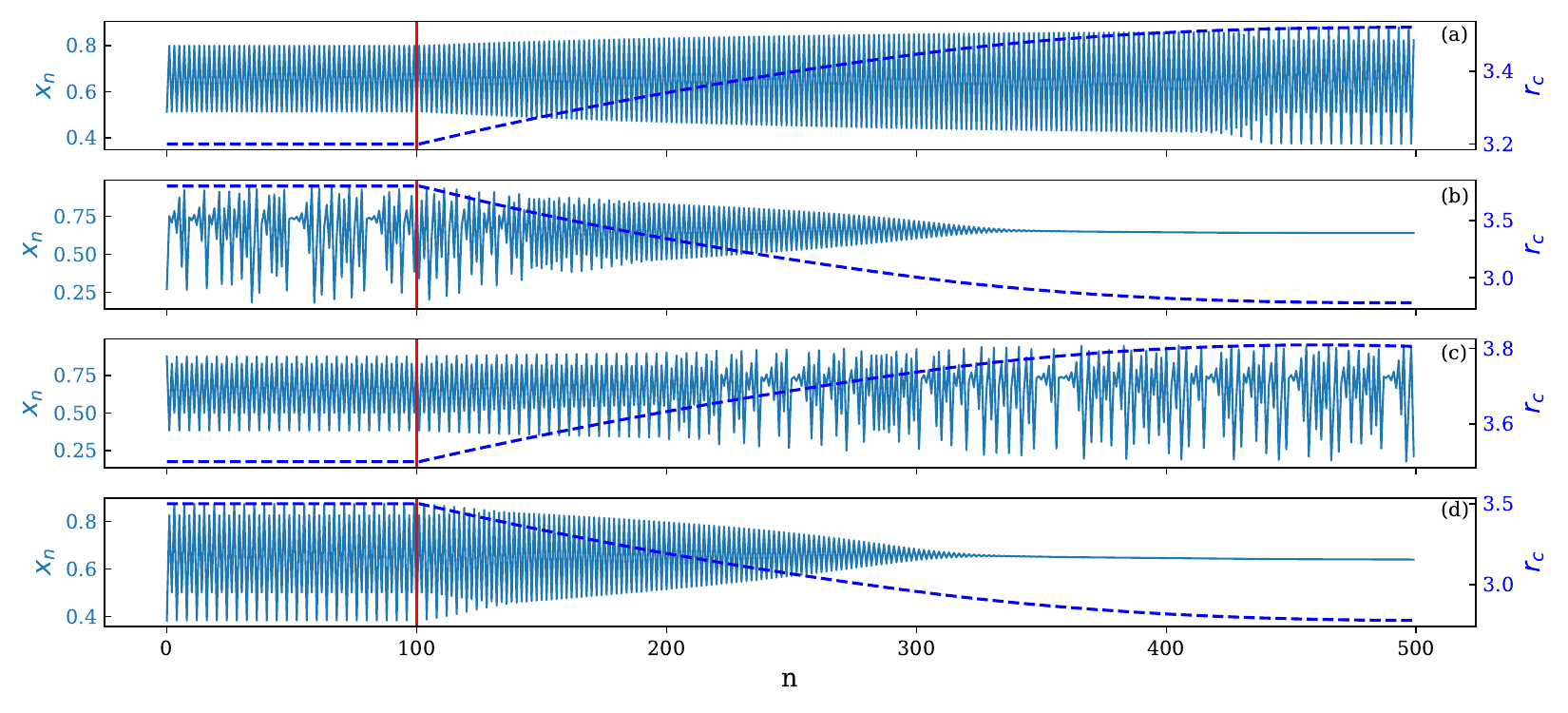} 
    \caption{Numerical results for control with parameter $r_c$ of the Logistic Map (\ref{eq:logm}). (a - d) show the time series of the system with control mechanism in action (blue lines) and the variation of the control parameter $r_c$ (dashed dark blue lines). The vertical red lines denote the start of application of the control signal. For all four rows the corresponding initial and target values of parameter of $r_c$ are given by $(3.2,3.5),~ (3.8,2.8),~(3.5,3.8)$, and $(3.5,2.8)$ respectively.}
    \label{fig:control_logm}
\end{figure*}

We employ a similar method of adding control signal ($r'$) to the actual parameter ($r$), resulting $r_c = r + r'$. Thus, the evolution of the control input $r'$ can be described as

\begin{equation}\label{eq:control_logm}
    r'_{n+1} = r'_n + \varepsilon_{logm}(r^* - r_p),
\end{equation}
where $r^*$ is the system parameter corresponding to target dynamics. $r_p (\approx r_c)$ is the prediction by the ESN from real-time measurement of the time series. For the numerical simulation results are presented in Figure \ref{fig:control_logm}, the training data samples are chosen for $r_c = 3.45,~3.50,$ and $3.55$. $r_p$ is evaluated by averaging last $200$ steps of ESN output. The stiffness of control $\varepsilon_{logm}$ is kept at $0.005$.

\section{\label{sec:expt}Control in Electronic Circuit}

\begin{figure*}
    \centering
    \includegraphics[width=\textwidth, keepaspectratio]{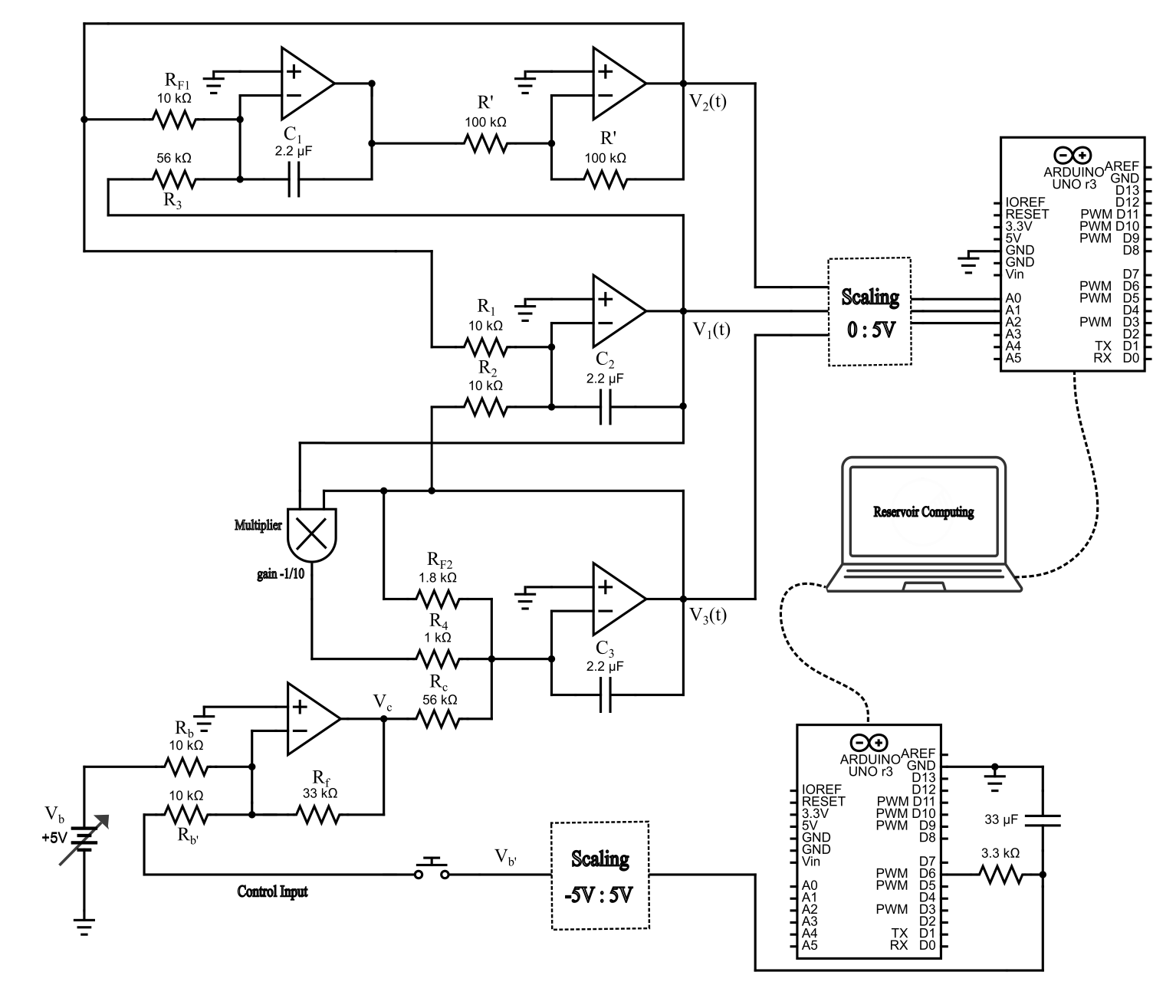}
    \caption{The control scheme implemented on the electronic circuit.}
    \label{fig:rossler-circuit}
\end{figure*}

To demonstrate the efficacy of our proposed approach, we implement the scheme to an experimentally constructed R\"{o}ssler oscillator. Figure \ref{fig:rossler-circuit} shows the electronic circuit diagram of the implementation.
The circuit is constructed using $TL082$ (JFET) operational amplifiers and $AD633AN$ four-quadrant analog multiplier integrated circuits (ICs). The operational amplifiers are powered by a $\pm 15V~DC$ supply, providing stable operation of the  R\"{o}ssler system dynamics.
Outputs of the integrators can be expressed by the following equations,

\begin{align}
V_1(t) &= -\frac{1}{C_1} \int \left( \frac{V_2}{R_1} + \frac{V_3}{R_2} \right) \, \mathrm{d}t, \nonumber 
\\[1em]
V_2(t) &= -\frac{1}{C_2} \int \left( -\frac{V_1}{R_3} - \frac{V_2}{R_{F1}} \right) \, \mathrm{d}t, \\[1em]
V_3(t) &= -\frac{1}{C_3} \int \left( -\frac{V_b}{R_c} - \frac{V_1V_3}{10R_4}+\frac{V_3}{R_{F2}}  \right) \, \mathrm{d}t.  \nonumber 
\end{align}

For this experiment, we set $R_1=R_2=R_3=R_4=R$  and $C_1=C_2=C_3=C$. Differentiation of the equations mentioned above with respect to $t$, and a subsequent scaling transformation $t' = t/RC$ yields the resulting voltage equations describing the dynamics of the circuit, which can be expressed as

\begin{align}
\frac{dV_1}{dt'} &=  -V_2 - V_3,  \nonumber \\[1em]
\frac{dV_2}{dt'}  &=   V_1  + \frac{R}{R_{F1}}V_2,  \\[1em] 
\frac{dV_3}{dt'}  &=  \frac{R}{R_c}V_c + V_1V_3 - \frac{R}{R_{F2}}V_3. \nonumber 
\end{align}

\begin{figure}[h]
    \centering
    \includegraphics[width=\linewidth]{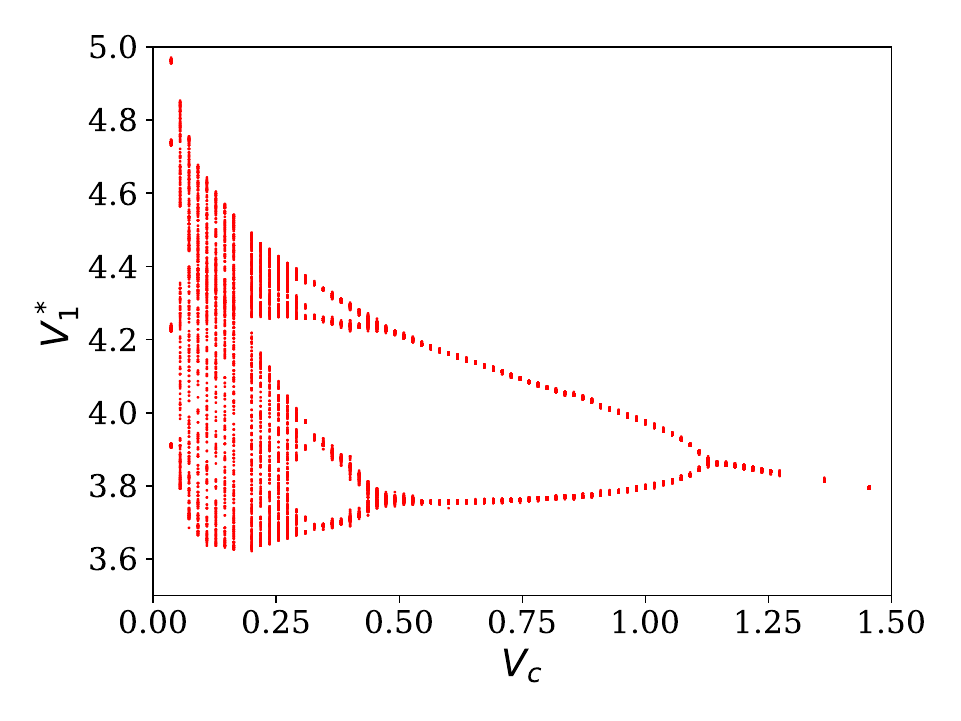}
    \caption{Bifurcation diagram of the dynamics of electronic circuit with varying $V_c$.}
    \label{fig:expt_bif}
\end{figure}

These equations resemble the dynamics of the R\"{o}ssler system given in equation (\ref{eq:control_ross}) in which $a\equiv \frac{R}{R_{F1}}$, $c\equiv \frac{R}{R_{F2}}$ and the control parameter $b_c \equiv \frac{R}{R_c}V_c$. We select $R = 10K\Omega$ and $C = 2.2nF$, then setting $R_{F1} = 56K\Omega$, $R_{F2} = 1.8K\Omega$, will result in  $a \approx 0.18$ and $c \approx 5.56$. Component values, including resistors and capacitors, are chosen carefully with minimal tolerance to maintain the precision required for the accurate replication of the dynamics of the R\"{o}ssler system. The change in dynamics with varying the voltage $V_c$ is plotted in the bifurcation diagram given in Figure \ref{fig:expt_bif}.




Here $V_c$ is the bifurcation parameter expressed as $V_c = V_b + V_{b'}$. $V_b$ is the actual (or initial) system parameter which should be stabilized against any external perturbation or to be tuned to achieve target dynamics. $V_{b'}$ is the control signal produced by RC to adaptively lead the voltage $V_c$ to a target state and maintain there, irrespective of the value of $V_b$.
Specifically, $V_b$ is introduced into the circuit via a resistor $R_b$ by utilizing a variable $5V$ DC power supply. The control signal $V_{b'}$ is inserted via a {\em control input} channel using a summing amplifier. The experimental Rössler system is initialized into any random initial attractor by fixing the variable power supply to an arbitrary value while ensuring that the control input key is maintained in the off position during the process. The proposed ML scheme is utilized to drive the randomly generated attractor towards the target attractor. This is accomplished by feeding the experimentally generated data to a well-trained RC model using an Arduino Uno micro-controller. Subsequently, the RC framework then predicts bifurcation parameter values with high accuracy. Upon predicting the parameter value, the machine computes the deviation of the predicted value from the target value and iteratively updates the new parameter value using the external control input equation 
\begin{equation}\label{eq:control_expt}
    V_{b'}[s+1] = V_{b'}[s] + \varepsilon_{expt}(V^{*} - V_p),
\end{equation}
where $V^{*}$ is the target value of the control parameter and $\varepsilon_{expt}$ is the stiffness constant. $V_p (\approx V_c)$ is the predicted value of $V_c$. When the {\em control input} channel is turned on, the updated parameter value $V_{b'}$ is fed into the circuit using a second micro-controller with proper scaling, ensuring that the circuit dynamics is now governed by the new updated value of $V_c$. The resulting time series is measured and fed into the reservoir, where the updated parameter value is predicted. Following equation (\ref{eq:control_expt}), the mechanism modifies the input signal, advancing the value of $V_c$ closer to the targeted value and reintroducing it to the circuit via the {\em input channel}. This process repeats autonomously, leading the system to the intended dynamics and making it robust against any perturbation with minimal errors. Notably, in this whole process the value of actual system parameter $V_b$ is insignificant. It just affects the time required to achieve the target state.

To assess the robustness of the proposed machine learning scheme, we intervened by deliberately altering the value of $V_b$ using the variable $5V$ DC power supply. This can be viewed as applying an external perturbation that is driving the system away from the desired dynamics. We observe that despite a significantly large deviation of the parameter value $V_b$, the mechanism successfully drives it back closer to the target value by employing steps that minimize the difference between the target and predicted values. A video this experimental realization is provided as supplementary material \cite{video_unp}.

\section{\label{sec:conc}Discussions and Conclusion}

We demonstrated the effectiveness of a machine learning-assisted adaptive control mechanism in dynamical systems through numerical simulation and experimental implementation. The control scheme is applicable for continuous-time flows as well as discrete-time maps. We showcase the results of control on the R\"{o}ssler system and the Logistic map. In both cases, we consider a wide range of attractors as target dynamics starting from arbitrarily different initial or perturbed dynamics. The proposed mechanism is also verified on a physical system by experimentally implementing it on a R\"{o}ssler circuit. Several earlier works with similar objectives specifically focus on chaos control, and not exactly on targeting a particular attractor. Our method relies on adjusting the system parameters, hence it can lead and maintain a system in the target state preserving all its dynamical properties in that particular state. One can target any already existing dynamics picked from the original bifurcation diagram of the system. Moreover, the control mechanism is only dependent on the measurement of the state variables; information about the original (or perturbed) value of the control parameter is not required. 

However, there are some points to note about this approach. One of the most significant assumptions on which our proposed method is based is that we have the flexibility to adjust the system parameter by external intervention. For general physical systems, identifying a method for this kind of intervention may be difficult. 
Moreover, most physical systems have nonlinear responses to a changes in system parameters. Thus, real-time control with the proposed approach needs to be slow with small values of stiffness of control. This leads to increased transient time to reach the target dynamics. Despite these restrictions, we believe that this work has a significant contribution to the application of machine learning tools to control dynamical systems.

\section*{Acknowledgment}
SM and KA acknowledge the financial support by JST Moonshot R\&D Grant Number JP-MJMS2021. KA also acknowledges Institute of AI and Beyond of UTokyo, JSPS KAKENHI Grant Number JP20H05921, and Cross-ministerial Strategic Innovation Promotion Program (SIP), the 3rd period of SIP, Grant Numbers JPJ012207, JPJ012425.
SC and MS acknowledge financial support from the Science and Engineering Research Board (SERB), Department of Science and Technology (DST), India (Grant No. CRG/2021/003301).

\bibliography{citations}

\begin{thebibliography}{30}
\bibitem [{vid()}]{video_unp}%
  \BibitemOpen
  \href@noop {} {}\bibinfo {note} {The video for the experimental demonstration
  of the control mechanism on electronic circuit is to be
  published.}\BibitemShut {Stop}%
\end{thebibliography}

\end{document}